# Multisession Longitudinal Dynamic MRI Incorporating Patient-Specific Prior Image Information Across Time

Jingjia Chen, Hersh Chandarana, Daniel K. Sodickson, Li Feng*

1. Bernard and Irene Schwartz Center for Biomedical Imaging, Department of Radiology, New York University Grossman School of Medicine, New York, New York, USA
2. Center for Advanced Imaging Innovation and Research (CAI2R), Department of Radiology, New York University Grossman School of Medicine, New York, New York, USA

Address correspondence to:

Li Feng, PhD

Center for Advanced Imaging Innovation and Research ($CAI^2R$)

New York University Grossman School of Medicine

660 First Avenue

New York, NY, USA 10016

Email: Li.Feng@nyulangone.org

## Abstract

Serial Magnetic Resonance Imaging (MRI) is common in clinical care, providing valuable longitudinal anatomical and motion information across imaging sessions. However, current reconstruction methods treat each session in isolation, ignoring this rich temporal context. In this work, we introduce a novel concept of longitudinal dynamic MRI, which incorporates patient-specific prior images to exploit temporal correlations across multiple imaging sessions. This framework enables progressive acceleration, reducing scan time as more imaging history becomes available. We demonstrate the feasibility using the 4D Golden-angle RAdial Sparse Parallel (GRASP) MRI, a state-of-the-art dynamic imaging technique. Multi-session GRASP datasets are concatenated into an extended dynamic series and reconstructed using a low-rank subspace algorithm. Across experiments, longitudinal reconstruction consistently outperforms single-session reconstruction in image quality while preserving true inter-session variations, including changes in anatomy, body contour, and imaging intervals. More broadly, this work suggests a new context-aware imaging paradigm in which the more we see a patient, the faster we can image.

## Introduction

Magnetic Resonance Imaging (MRI) is a versatile and powerful imaging modality that provides high-resolution, multi-contrast information about tissue anatomy and function, enabling non-invasive diagnosis, disease monitoring, and treatment planning. However, despite the many advantages of MRI, its inherently slow imaging speed remains a major limitation compared to other modalities.

Over the past two decades, a variety of rapid imaging techniques have been developed to accelerate MRI acquisition, including parallel imaging[1–3], sparsity-based reconstruction[4,5], and, more recently, deep learning[6–9] approaches. These advances have all demonstrated a profound clinical impact and have been well-established or increasingly adopted in routine imaging workflows. They have also enabled highly accelerated dynamic MRI acquisitions in free-breathing body applications, reducing or even eliminating the need for breath holds in many cases. However, most existing MRI reconstruction methods today are designed to generate clean images from individual scans in single imaging sessions, without considering prior image information often available from the same patient. This represents a missed opportunity, as many patients need to undergo repeated MRI exams for longitudinal assessment, such as monitoring disease progression, evaluating treatment response, or guiding therapy planning over time. These repeated MRI scans, referred to here as longitudinal imaging, contain substantial shared information across imaging sessions that could potentially be leveraged to improve image reconstruction and accelerate the imaging process still further.

In fact, exploiting prior image information over short time scales to improve reconstruction quality is a well-established concept in the field of MRI. In state-of-the-art dynamic MRI reconstruction, image data from adjacent temporal frames acquired seconds to minutes earlier is commonly used to exploit temporal redundancy and enable higher acceleration rates for each dynamic frame[10–15]. For example, Golden-angle RAdial Sparse Parallel (GRASP) MRI[16–18] is a well-recognized dynamic free-breathing imaging technique that combines compressed sensing and parallel imaging with golden-angle radial sampling to leverage time information for higher acceleration rates. Similarly, joint multi-contrast reconstruction[19–23] is another example of utilizing short-time scale correlations to aid image reconstruction, where an image with one contrast (e.g., T1-weighted) acquired a few minutes earlier can assist in reconstructing a subsequently acquired image with a different contrast (e.g., T2-weighted) from the same patient.

However, despite this long history of using intra-session, short-time-scale temporal information, very little attention has been given to leveraging shared information across multiple imaging sessions to exploit longitudinal temporal redundancy over a longer time scale. This raises a compelling question: can we extend our dynamic reconstruction strategies to incorporate patient-specific prior images acquired over longer intervals, like days, weeks, or even months, to improve reconstruction quality and enable higher acceleration rates? Variations between imaging sessions in the same patient are generally limited to changes in scan position, physiological state, disease progression, or treatment effects. These variations can be treated as dynamic changes, analogous to how respiratory or cardiac motion is handled in conventional dynamic MRI reconstruction. Such an approach has the potential to enable higher acceleration rates when multiple imaging sessions are reconstructed jointly, in contrast to traditional methods that rely solely on information from individual sessions in isolation.

In this study, we propose a novel concept for longitudinal dynamic MRI reconstruction that incorporates patient-specific prior image information across extended time scales. More specifically, we aim to demonstrate this idea by developing a longitudinal dynamic MRI framework based on the GRASP imaging technique to enable free-breathing, time-resolved 4D MRI that leverages temporal correlations across imaging sessions through multi-session joint reconstruction. The feasibility, performance, and robustness of this framework were evaluated through a series of proof-of-concept experiments. Our main hypothesis is that incorporating patient-specific longitudinal data from prior MRI scans can enable progressive acceleration of data acquisition beyond what can be achieved when each imaging session is reconstructed independently. In other words, we expect that the more scans a patient undergoes, the greater the acceleration that can be achieved during the current scan. Such an imaging approach is expected to have important clinical implications for scenarios involving repeated dynamic imaging, such as MRI-guided radiation therapy (MRgRT), where 4D MRI is increasingly being adopted for visualizing and tracking the movement of abdominal or thoracic tumors over the course of treatment[24–29]. To the best of our knowledge, this is the first study to explore longitudinal dynamic image reconstruction leveraging shared information across multiple imaging sessions. We believe this opens a new avenue for rethinking how dynamic imaging can be performed in longitudinal clinical settings, with strong potential to improve both reconstruction quality and imaging efficiency across a wide range of clinical applications.

## Results

The proposed longitudinal dynamic imaging framework is implemented based on the GRASP MRI technique[16], which was originally developed for rapid dynamic volumetric MRI exploiting short-time scale information within a single imaging session. The use of golden-angle radial sampling in GRASP MRI enables continuous data acquisition during free breathing and allows for flexible reconstruction of dynamic images with varying temporal resolutions from the same dataset. Over successive iterations, GRASP has advanced into a highly accelerated framework, with the latest version used in this work [30,31] achieving time-resolved 4D MRI with sub-second temporal resolution. In all reconstruction tasks, two consecutive golden-angle radial spokes along with one 2D navigator spoke were combined into one dynamic frame (**Figure 1**), and 3D volumes were constructed from multiple 2D frames, resulting in a temporal resolution of 282 ms per volume.

When 4D GRASP imaging sessions are repeatedly performed on the same patient, all available 4D GRASP datasets can be concatenated into a single *extended time-resolved dynamic series* for joint reconstruction, as illustrated in **Figure 1**. In this work, this reconstruction strategy is referred to as longitudinal multi-session reconstruction or longitudinal 4D GRASP MRI. While changes in patient positioning or body shape may occur across sessions, these variations, as previously mentioned in the Introduction, can be treated as extended dynamic changes to be captured and recovered during the reconstruction process, similar to how respiratory and cardiac motion are addressed in standard dynamic MRI reconstruction. In addition, an optional 3D rigid registration step can be applied to improve inter-session alignment. This registration can be performed on a time-averaged 3D image volume reconstructed separately from each session and then applied to the corresponding radial k-space data.

In this study, a total of six experiments were conducted to evaluate the performance of longitudinal 4D GRASP MRI and to address key questions related to this new imaging framework, as summarized below.

**Experiment 1: To assess whether incorporating subject-specific prior imaging sessions enables higher acceleration rates while maintaining image quality.**

This first experiment was designed to address the central question of this study: whether incorporating prior imaging data from the same subject enables higher acceleration rates without compromising image quality. This experiment was performed on a healthy volunteer, with the second and third imaging sessions performed 7 and 12 days after the initial session, respectively. In each imaging session, 1000 golden-angle rotated spokes were acquired for every slice, which are expected to provide sufficient temporal correlations to ensure reconstruction of high-quality time-resolved 4D GRASP images for each individual session separately[31,32]. Dynamic images reconstructed using all 1000 spokes serve as the references for image quality comparison.

Longitudinal 4D GRASP reconstruction task was performed, including three imaging sessions, with 500 spokes from the first session, 300 spokes from the second session, and 200 spokes from the third session for joint reconstruction. Different spokes were chosen from each session to ensure non-repeating sampling trajectories for longitudinal 4D GRASP reconstruction, as illustrated in **Supplementary Figure S1**. Single-session 4D GRASP reconstruction without utilizing longitudinal correlations was performed using matching spokes for comparison. In addition, reference 4D images for each imaging session were reconstructed using all 1000 s pokes from that session, and they serve as the ground truth for evaluation. The single-session and reference reconstructions were performed in all subsequent experiments and will not be mentioned further in each experiment below.

**Figure 2** compares longitudinal 4D GRASP reconstruction with standard single-session 4D GRASP reconstruction. The images are snapshots of one dynamic frame from each imaging session. Note that a smaller number of total spokes/temporal frames results in lower temporal correlations, which degrades the reconstruction quality in single-session 4D GRASP reconstruction. Longitudinal 4D GRASP reconstruction demonstrates superior visual image quality compared to single-session reconstruction with a matching number of spokes. The improvement is confirmed by the quantitative metrics shown above the images. Corresponding cine movies for this comparison are provided in **Video 1** (Supplementary Materials).

**Experiment 2: To assess whether longitudinal reconstruction maintains good image quality and accuracy in the presence of pathological lesions.**

In this experiment, we aimed to assess the performance of longitudinal 4D GRASP reconstruction on one subject with a lesion identified in the liver. The subject underwent the second and third imaging sessions 10 and 18 days after the initial session, respectively. All imaging procedures and assessments in this experiment (and in all subsequent experiments) followed the same protocol described in Experiment 1.

**Figure 3** compares longitudinal 4D GRASP reconstruction with single-session reconstruction in a subject with a liver lesion, indicated by the red arrows. Longitudinal reconstruction improves the overall image quality compared to single-session reconstruction with a matching number of spokes, particularly in the third session with the fewest spokes. Longitudinal reconstruction also enables a better delineation of the lesion in all imaging sessions. These improvements are further confirmed by the quantitative metrics shown above the images. Corresponding cine movies for this comparison are provided in **Video 2** (Supplementary Materials).

**Experiment 3: To assess whether longitudinal reconstruction remains effective in the presence of significant changes across sessions.**

This experiment aimed to evaluate whether major changes occurring between imaging sessions could be preserved during longitudinal 4D GRASP reconstruction. Imaging was performed on one subject with a body mass index (BMI) of 31 and correspondingly high body fat content. The subject underwent the second and third imaging sessions 24 and 27 days after the initial session, respectively. To introduce variations across sessions, the sequence timing for fat suppression was intentionally modified for the MRI scan during the second session. This resulted in reduced fat suppression performance and thus increased fat signal in the images of this session compared to the first and third sessions.

**Figure 4** evaluates the performance of longitudinal 4D GRASP reconstruction in the presence of these inter-session changes. As indicated by the red arrow in the figure, images from the second session show increased residual fat signal. In this comparison, longitudinal reconstruction outperforms single-session reconstruction while successfully preserving the differences in fat signal. The most significant improvements in image quality are observed in the second and third sessions, which had fewer spokes. These results demonstrate the robustness of our longitudinal reconstruction approach in handling inter-session changes while maintaining high image quality. Corresponding cine movies for this comparison are provided in **Video 3** (Supplementary Materials).

**Experiment 4: To assess whether longitudinal reconstruction maintains performance with large inter-session gaps.**

In this experiment, we evaluated the performance of longitudinal 4D GRASP reconstruction on one subject who underwent the second and third imaging sessions 29 and 274 days after the initial session, respectively. Given the extended time interval, we expected that changes in body habitus might have occurred between imaging sessions and aimed to assess whether the longitudinal reconstruction could preserve such variations over a longer inter-session gap.

**Figure 5** presents the results of longitudinal 4D GRASP reconstruction, where the interval between imaging sessions was significantly extended. As shown in the reference images, notable changes in body contour are observed between the first two sessions and the third session, which could be due to changes in body habitus during the large inter-session interval. Despite these significant changes, the longitudinal 4D GRASP reconstruction effectively managed the extended time intervals, yielding improved image quality compared to single-session 4D GRASP reconstruction. This experiment demonstrates the robustness of our longitudinal reconstruction approach in handling substantial variations not only in imaging strategy but also in body structure across imaging sessions. Corresponding cine movies for this comparison are provided in **Video 4** (Supplementary Materials).

**Experiment 5: To assess whether subject-specific prior information is more valuable than generic population-based information for longitudinal reconstruction.**

A key assumption of longitudinal 4D reconstruction is that datasets from different imaging sessions are acquired from the same subject. One may ask: what if datasets from different subjects are instead combined for joint reconstruction? Could this population-based approach achieve similar performance to subject-specific longitudinal reconstruction? To address this question, we compared joint 4D GRASP reconstruction using three imaging sessions from the same subject versus three sessions from three different subjects. For the joint reconstruction across different subjects, 500 spokes were taken from the first subject, 300 from the second, and 200 from the third, forming a pseudo-longitudinal dataset. As with subject-specific longitudinal 4D GRASP reconstruction, spokes were selected to ensure non-repeating sampling trajectories.

**Figure 6** compares longitudinal 4D GRASP reconstruction using three imaging sessions of the same subject versus three sessions of different subjects. While the difference in reconstructed image quality is minimal in the first session (500 spokes), longitudinal 4D GRASP reconstruction achieves better image quality in sessions 2 and 3, where images from the same subject exhibit higher structural clarity and fewer artifacts. In contrast, concatenating data from different subjects introduces noticeable artifacts and blurring. This indicates that subject-specific prior information is, in fact, being leveraged in longitudinal 4D GRASP reconstruction.

**Experiment 6: To assess whether the advantages of longitudinal as compared with single-session reconstruction were seen consistently across multiple subjects.**

Longitudinal 4D GRASP reconstruction and single-session 4D GRASP reconstruction, using a matched number of spokes for each session, were compared across all eight subjects by calculating the Structural Similarity Index Measure (SSIM) and Normalized Root Mean Squared Error (NRMSE). Ground truth images for these quantitative metrics were generated as described before, using all 1,000 spokes. The metrics were computed for all corresponding frames within each session, and statistical differences between methods were evaluated using one-tailed paired Student's *t*-tests.

**Figure 7** summarizes the quantitative results for longitudinal 4D GRASP reconstruction across all the subjects, with asterisks denoting statistical significance *: $p<0.05$, and **: $p<0.01$. Detailed p-values are summarized in the table below. Consistent with the above findings in each experiment, longitudinal reconstruction demonstrates the best performance across both metrics, while single-session 4D GRASP reconstruction yields the lowest overall image quality compared to longitudinal approaches.

## Discussion

Image reconstruction has long been one of the most active research areas in the field of MRI. However, despite decades of extensive research focused on accelerated image acquisition, little attention has been given to exploiting longitudinal information, even though such information is readily available from repeated MRI scans in current clinical practice. In this study, we demonstrate a novel framework for longitudinal dynamic reconstruction that incorporates patient-specific prior image information across time. At the time of this work, only a few studies[33,34] had demonstrated the feasibility of static image reconstruction using patient-specific prior information, and to the best of our knowledge, no studies had explored the use of longitudinal information in dynamic image reconstruction.

### Improvement of longitudinal reconstruction over standard reconstruction

In this work, the performance of longitudinal dynamic image reconstruction was demonstrated using the GRASP MRI technique previously developed by our team. In particular, we showcased time-resolved longitudinal 4D MRI, which naturally enables the concatenation of longitudinal data across time and incorporates multiple imaging sessions. Our technique employs a low-rank subspace-based image reconstruction scheme to achieve high-quality, time-resolved longitudinal 4D MRI with high temporal resolution. As detailed in the Methods section below, our low-rank subspace-based reconstruction algorithm is implemented in two steps: a temporal basis is first estimated to represent the dynamic images in a low-dimensional subspace. Iterative reconstruction is then performed in this subspace to remove undersampling artifacts and generate clean images. Compared to standard single-session dynamic MRI reconstruction, our longitudinal approach offers two key advantages to achieve higher acceleration rates without compromising image quality. First, incorporating patient-specific prior information allows for the estimation of a more accurate temporal basis, which is essential to our reconstruction algorithm. As demonstrated in **Supplementary Figure S2**, using a jointly estimated temporal basis already improves image reconstruction for single-session reconstruction. Second, concatenating dynamic data from multiple imaging sessions increases temporal correlations, which can be exploited during iterative reconstruction. This enables improved reconstruction quality at higher acceleration rates compared to standard single-session reconstruction. Together, these improvements enhance overall reconstruction quality while allowing for progressively reduced scan time in each imaging session.

**Robustness against inter-session variations**

We have shown that our technique is robust to inter-session variation, which was intentionally introduced in our experiments with different fat suppression settings in Experiment 3. Despite the longitudinal changes, our method was able to preserve structural details across sessions successfully, yielding higher image quality compared to single-session reconstruction. In Experiment 4, which involved extended inter-session intervals and noticeable variations in body contour, our approach remained effective, even in the presence of substantial temporal gaps and anatomical changes within the same subject.

By default, the longitudinal 4D GRASP reconstruction incorporates a pre-processing step of rigid registration for image alignment. As demonstrated in **Supplementary Figure S3**, while this alignment step does improve reconstruction quality, the improvement from this pre-processing step is relatively modest. While the liver region largely can be aligned through registration, the physiological differences in the stomach and its relative position to the spleen cannot be matched through rigid registration. Importantly, despite these non-systematic variations, our longitudinal reconstruction still preserves such session-specific physiological changes without introducing contrast leakage or blurring. This likely stems from the inherent ability of our method to account for temporal variations by treating them as dynamic changes, akin to how respiratory and cardiac motion are handled in standard dynamic MRI reconstruction. These findings highlight the potential of longitudinal 4D GRASP reconstruction for applications where imaging sessions may be spaced over longer durations. Our technique is expected to manage inter-session differences without this process.

**Importance of Subject-Specific Prior Image Information**

Our technique relies on longitudinal imaging of the same subject to fully leverage cross-session dynamic correlations. When datasets from different subjects are concatenated to create pseudo-longitudinal sessions, the method no longer benefits from such accumulated information. This is because anatomical structures, organ motion patterns, and image intensity distributions can vary significantly across individuals and thus disrupt the underlying temporal sparsity/correlations that our approach exploits. As demonstrated in Experiment 5, incorporating data from different subjects leads to remaining artifacts and blurring that cannot be completely removed, which highlights the importance of maintaining a consistent anatomical structure for effective longitudinal reconstruction. However, we do note that joint reconstruction combining data from different subjects still shows a slight improvement in image quality compared to standard separate single-session reconstruction. This is likely because all datasets, despite being from different subjects, still originate from generally similar abdominal anatomy. This results in some degree of shared anatomical and dynamic characteristics, even though the correlations are not as strong as those from the same subject.

### Potential Clinical Applications

One of the most immediate clinical applications of our proposed technique is in scenarios where repeated dynamic MRI scans are required. A prominent example is MRgRT using MRI-Linac systems, which integrate MRI with a linear accelerator (Linac) to enable more precise radiation delivery. In MRgRT, 4D MRI has increasingly been used for treatment planning in moving organs such as the lungs and liver. However, 4D MRI typically requires longer acquisition times compared to standard 2D or 3D imaging, making it less feasible for daily adaptive workflows. As a result, 4D MRI is often limited to initial treatment planning rather than routine use throughout the course of therapy, although this is preferred. The proposed longitudinal 4D MRI technique offers a promising solution to this challenge by enabling faster, higher-quality 4D imaging across fractions, particularly when multi-contrast 4D MRI is required. Moreover, time-resolved 4D MRI, as implemented in our imaging framework, enables more accurate tracking of respiratory variations over time, including motion drift and irregular breathing[32]. Beyond radiotherapy, another application of our longitudinal dynamic MRI framework is in repeated dynamic contrast-enhanced (DCE) MRI, which is frequently used to monitor disease progression in cancer patients.

### Limitations and Future Work

While our study demonstrates the feasibility of longitudinal 4D GRASP reconstruction, several limitations warrant discussion. First, this is an initial proof-of-concept study with a limited number of subjects. Future studies with a larger cohort of patients across diverse clinical conditions will be essential to further validate the performance of this technique. Second, due to practical challenges of recruitment for this proof-of-concept study, we did not include patients with new lesions appearing between distinct imaging sessions. However, we did validate the performance of the technique both with consistent lesions and in the presence of assorted inter-session variations, and our results provide confidence that the method would not miss important pathological changes in extended longitudinal studies. Third, our current study focuses only on T1-weighted dynamic imaging to evaluate the proposed longitudinal reconstruction approach. Future work will extend this framework to other contrast mechanisms, such as T2-weighted imaging, which would allow for a more comprehensive evaluation and enhance clinical utility in routine practice. Fourth, our longitudinal 4D GRASP reconstruction is implemented using an iterative reconstruction algorithm in this proof-of-concept study. We expect that deep learning-based reconstruction methods will further improve performance and reduce reconstruction time.

## Conclusion

In summary, we have introduced a novel concept of longitudinal dynamic image reconstruction that leverages cross-session dynamic correlations to improve image quality and accelerate data acquisition. Our approach demonstrated robustness to inter-session changes, varying time intervals, and differences in body contour, while effectively preserving structural details, including lesions. Although further validation with larger patient cohorts is pending, our initial findings highlight the potential of longitudinal dynamic imaging for applications that require repeated time-resolved or real-time imaging over extended periods. These findings also suggest that it is possible, using modern reconstruction methods, to place data from any given imaging examination is a broader subject-specific context, with resulting benefits for efficiency, efficacy, and accessibility.

## Methods

### Human Subjects and Data Acquisition

Eight human subjects (4 females and 4 males, mean age = 44.4 ± 15.8 years) were recruited for the various MRI experiments in this study, which was HIPAA-compliant and approved by the local Institutional Review Board (IRB). Written informed consent was obtained from all participants before MR scans. Each subject underwent three separate MRI scans on different days on a 3T clinical MRI scanner (Siemens

MAGNETOM Prisma, Erlangen, Germany) to emulate three longitudinal imaging sessions.

Imaging in all three sessions was performed using a 3D fat-suppressed radial sequence with a modified stack-of-stars sampling trajectory that incorporated additional 2D navigators[30–32,35] (see **Figure 1**). Specifically, one 2D navigator with a consistent acquisition angle of zero degrees was acquired after every two regular golden-angle spokes. As a result, each dataset included a total of 500 2D navigators and 1,000 golden-angle radial spokes in each imaging slice. The total scan time for each session was 2 minutes and 21 seconds. Additional imaging parameters were as follows: field of view (FOV) = 360x360 $mm^2$, matrix size = 256 x 256, in-plane spatial resolution = 1.4 x 1.4 $mm^2$, slice thickness = 6 mm, repetition time (TR) = 2.51 ms, echo time (TE) = 1.16 ms, flip angle (FA) = $10^o$, number of slices = 40, and slice partial Fourier = 75%.

**Time-Resolved 4D GRASP MRI**

The proposed longitudinal dynamic imaging framework is based on the GRASP MRI technique and enables time-resolved 4D MRI across multiple sessions. In this section, we first describe the standard time-resolved 4D GRASP MRI method. In the subsections to follow, we describe the extension of this method to incorporate longitudinal dynamic image information and explain how this enhances reconstruction performance.

The GRASP technique was originally developed for rapid, free-breathing dynamic MRI[16]. By combining golden-angle radial sampling with multicoil compressed sensing reconstruction, GRASP enables continuous data acquisition during free breathing, allowing for flexible reconstruction of dynamic images with varying temporal resolutions from the same dataset. The GRASP technique has evolved through multiple generations, with the latest version, used in this work, enabling time-resolved 4D MRI with sub-second temporal resolution[30,31]. We will refer to this technique henceforward as 4D GRASP MRI.

4D GRASP MRI acquisition employs a new sampling trajectory called navi-stack-of-star sampling. This approach is a variation of the original stack-of-stars method, in which a 2D navigator is periodically acquired, as shown in **Figure 1**. In the simplest implementation, these 2D navigators can be obtained as radial stacks acquired at a fixed angle (e.g., zero degrees)[30,32,35]. From these zero-degree stacks, coronal projections of the imaging volume can be generated with a 2D fast Fourier transform (FFT), which can be used to track respiration and bulk motion, as well as to calculate a temporal basis for low-rank subspace-based reconstruction, as described below.

Image reconstruction in 4D GRASP MRI is performed in a slice-by-slice manner after applying an FFT along the slice dimension to separate slice encoding in stack-of-stars data. For each image slice, low rank subspace-based 4D MRI reconstruction is performed by solving the following optimization problem:

$$\tilde{V}_K = \arg\min_{V_K} \frac{1}{2} \left\| E(V_K U_K') - \sqrt{W} y \right\|_2^2 + \lambda_t \| S_t(V_K U_K') \|_1 + \lambda_s \| S_s V_K \|_1 \quad [1]$$

where the dynamic images to be reconstructed with a matrix size of $N \times N$ and $T$ temporal frames are denoted as $m \in \mathbb{C}^{N^2 \times T}$ , and $y$ represents the acquired multicoil radial k-space data, which is shifted onto a Cartesian grid in a pre-processing step using the self-calibrating GeneRalized Autocalibrating Partial Parallel Acquisition (GRAPPA) Operator Gridding (GROG) approach[36,37]. $E$ is a multi-coil encoding operator incorporating the FFT operation, coil sensitivity maps, the underlying k-space undersampling pattern, and $\sqrt{W}$, a density weighting matrix estimated from the GROG pre-processing as described previously[36]. To enforce a low-rank subspace constraint on dynamic images, $m$ is represented as $V_K U_K'$, where $U \in \mathbb{R}^{T \times T}$ is the temporal basis pre-estimated from the 2D navigators using principal component analysis (PCA), $U_K \in \mathbb{R}^{T \times K}$ represents the $K$ (where $K \ll T$) dominant basis functions in $U$ for constructing the low-rank subspace, and $V_K \in \mathbb{C}^{N^2 \times K}$ represents the coefficients (also known as the spatial basis) associated with $U_K$ to be reconstructed during the optimization. $S_t$ and $S_s$ denote finite difference operators applied along the temporal dimension of $V_K U_K'$ and the spatial dimension of $V_K$ to enforce a temporal total variation (TV) constraint and a spatial TV constraint, respectively, with regularization parameters $\lambda_t$ and $\lambda_s$. After reconstructing $V_K$, the dynamic images $m$ can then be generated as $V_K U_K'$.

Depending on the frequency of 2D navigator acquisition, $m$ can be reconstructed with different temporal resolutions. For example, when each 2D navigator is acquired every two golden-angle rotations, as shown in **Figure 1**, each temporal frame in $m$ can be reconstructed using two spokes adjacent to each navigator in each slice. This approach enables a sub-second temporal resolution for time-resolved 4D MRI[17,30–32].

### Considerations for Standard Single-Session 4D GRASP MRI Reconstruction

The reconstruction problem formulated in Equation [1] primarily relies on temporal constraints through the low-rank subspace and the temporal total variation (TV) regularization. Although a spatial TV constraint is also included, its regularization weight ($\lambda_s$) is typically much smaller than the temporal regularization weight $\lambda_t$. Therefore, although standard 4D GRASP MRI has demonstrated the ability to achieve high temporal resolution (< 500 ms per 3D volume)[17,30–32], its successful implementation requires a sufficient number of temporal data points/frames to ensure (a) accurate estimation of a temporal basis for guiding low-rank-based image reconstruction, and (b) a clean DC component in the low-rank subspace. Here, the DC component in the low-rank subspace represents the average of all temporal frames. For the first point, reducing the number of temporal frames or correlations can lead to errors in estimating the temporal basis, thus resulting in degraded reconstruction performance (see **Supplementary Figure S2** for further details). For the second point, insufficient temporal frames or correlations can lead to a DC component with residual undersampling artifacts (see **Supplementary Figure S4** for an example). Since 4D GRASP MRI primarily relies on temporal regularization, the reconstruction is less effective at removing residual spatial artifacts in this scenario. As a result, standard single-session 4D GRASP MRI reconstruction requires adequate scan time to ensure sufficient temporal correlations, despite its ability to achieve high temporal resolution.

### Longitudinal Multi-Session 4D GRASP MRI

If repeated imaging sessions are performed on the same patient and 4D GRASP MRI is acquired in each session, all available 4D GRASP datasets can be concatenated into a single extended time-resolved dynamic series for joint reconstruction, as illustrated in **Figure 1**. In this work, this reconstruction strategy is referred to as longitudinal 4D GRASP reconstruction or longitudinal 4D GRASP MRI. While changes in patient positioning or body shape can occur across sessions, these variations can be treated as dynamic changes to be resolved during the reconstruction process, similar to how respiratory and cardiac motion are addressed in standard dynamic MRI reconstruction.

The selection of segments from the three sessions for both single-session and longitudinal 4D GRASP reconstructions is shown in **Figure 2**. For longitudinal 4D GRASP reconstruction, a subset of spokes was selected from each imaging session, and the selected spokes were then concatenated for joint reconstruction. The spokes from each session were chosen to ensure non-repeating sampling trajectories, as shown in **Supplementary Figure S1**. Regularization parameters for standard 4D GRASP reconstruction and longitudinal 4D GRASP reconstruction were optimized empirically on the acquired datasets separately to ensure temporal fidelity, and they are fixed for all the subject datasets.

The specific implementation of longitudinal 4D GRASP reconstruction involves the following steps. First, a single averaged 3D image is reconstructed from each session by combining all data within that session. These averaged 3D images are used to compute a 3D rigid registration across different sessions, which is subsequently applied to the corresponding radial k-space data in each imaging session to improve data alignment. To access the impact of the 3D rigid registration step, we additionally performed **Supplementary Experiment 1** comparing the longitudinal reconstruction with and without the registration step, as shown in **Supplementary Figure S4.** Results show that our reconstruction algorithm inherently compensates for intersession misalignment, with only minor improvements observed when the alignment step is included. Second, the 2D navigators from all imaging sessions are concatenated to estimate a joint temporal basis, $U_K$, for the concatenated datasets. As shown in **Supplementary Figure S2**, a more accurate temporal basis can be obtained by estimating from concatenated dynamic datasets when the scan time is reduced in each session. Third, the radial k-space data from all sessions are also concatenated into a single extended time-resolved dynamic series for joint reconstruction, following the optimization outlined in Equation [1] to reconstruct $V_K$, the spatial basis corresponding to the concatenated datasets. Finally, the extended dynamic images concatenated from all imaging sessions are generated as $V_K U_K'$. To ensure maximum temporal incoherence, radial data from different sessions can be acquired with non-repeating golden-angle rotation angles[38]. In order to assess the impact of sampling trajectory selection, additional experiments were performed **Supplementary Experiment 2** using repeating sampling trajectories and results were compared, as shown in **Supplementary Figure S5 and S6**.

By concatenating multi-session dynamic data together for joint reconstruction, our proposed longitudinal dynamic imaging reconstruction approach ensures (a) more accurate estimation of a temporal basis due to increased temporal correlations and (b) an artifact-free DC component in the low-rank subspace with multi-session data, even with reduced scan durations for some sessions. This reconstruction strategy enables progressive acceleration of data acquisition in longitudinal MRI as imaging sessions accumulate.

**4D MRI reconstruction times**

The computation time for the iterative reconstruction steps was compared across different methods. The average reconstruction time for longitudinal 4D GRASP reconstruction was 112.0±8.6 seconds per image slice, using a total of 1000 spokes concatenated from different imaging sessions (500 spokes from the first session, 300 spokes from the second session, and 200 spokes from the third session). For single-session 4D GRASP reconstruction, the average reconstruction time was 31.0±0.9 seconds per slice for the first session (500 spokes), 18.0±0.3 seconds for the second session (300 spokes), and 12.8±0.2 seconds for the third session (200 spokes). For the reference 4D GRASP reconstruction, which used all 1000 spokes from each session separately, the average reconstruction time was 107.8±0.94 seconds per slice. Longitudinal 4D GRASP reconstruction requires more time than single-session reconstruction with a reduced number of spokes, but its reconstruction time is comparable to reference 4D GRASP reconstruction using the same total number of spokes.

## Acknowledgments

This work was supported in part by the NIH (R01EB030549, R01EB031083, and P41EB017183) and was performed under the rubric of the Center for Advanced Imaging Innovation and Research (CAI$^2$R), an NIBIB National Center for Biomedical Imaging and Bioengineering. The authors would like to thank Ding Xia and Kai Tobias Block for their support in data acquisition, and Mary Bruno for her assistance with subject recruitment.

## Funding Support

This work was supported in part by the U.S. National Institutes of Health (grant numbers: R01EB030549, R01EB031083, and P41EB017183).

## Authors Contributions

J.C., D.S.K and L.F. contributed to the development of the reconstruction. J.C. and L.F. conducted all the MRI experiments. J.C. performed image reconstruction experiments and analysis. H.C., L.F. and D.S.K advised on the design of the experiments and provided supervision. J.C. wrote the initial draft of the manuscript. H.C., L.F. and D.S.K reviewed and revised the manuscript.

## Competing interests

The authors declare no competing interests.

## Disclosure

Li Feng, Hersh Chandarana, and Daniel Sodickson are co-inventors of a patent portfolio on the GRASP MRI technique.

## Figures

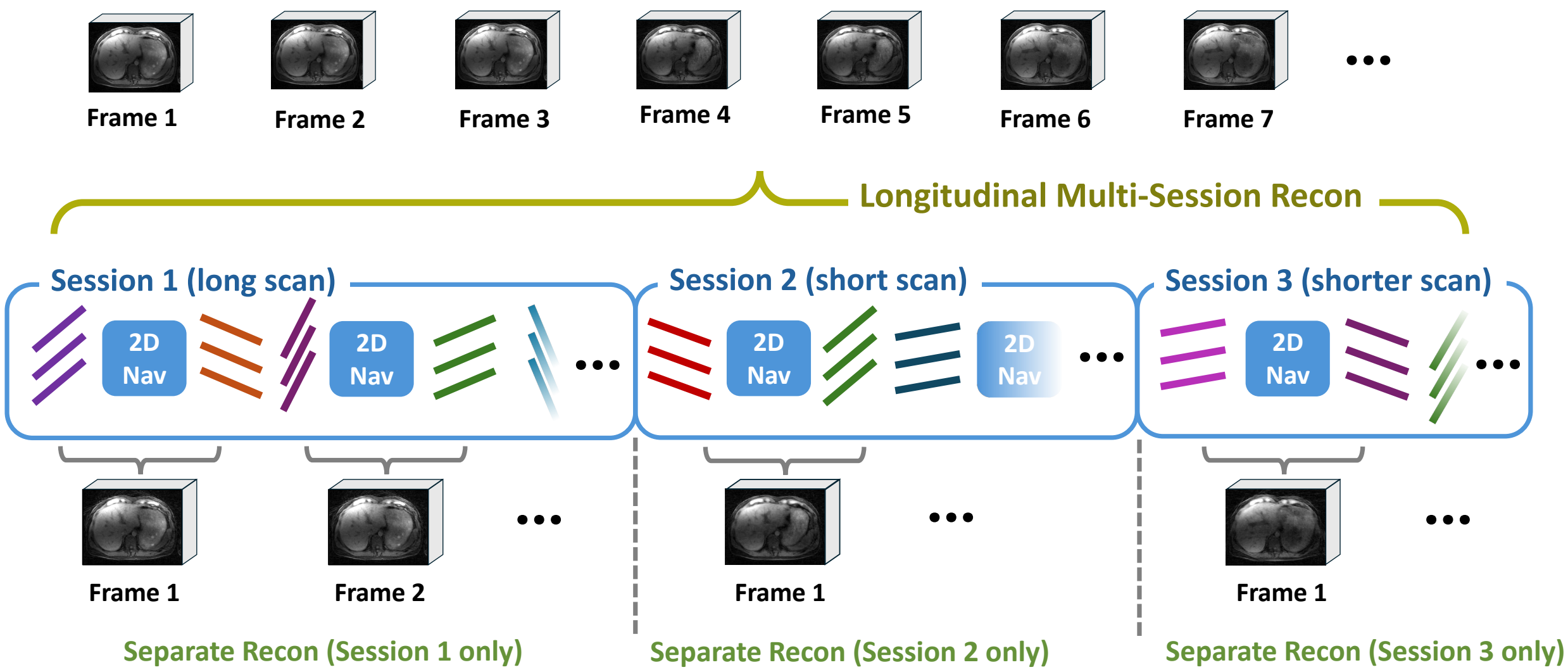

**Figure 1.** Schematic illustration of the proposed multi-session longitudinal 4D GRASP MRI reconstruction. Dynamic MRI data is acquired using navi-stack-of-stars sampling, which incorporates frequent 2D navigators into the golden angle radial stacks. The rotating angle follows a golden-angle increment that continues across sessions. Datasets from different sessions are concatenated along the temporal dimension to reconstruct dynamic images. Session 1 has the longest acquisition duration, while the follow-up sessions are shorter, demonstrating the progressive acceleration achieved through longitudinal multi-session reconstruction.

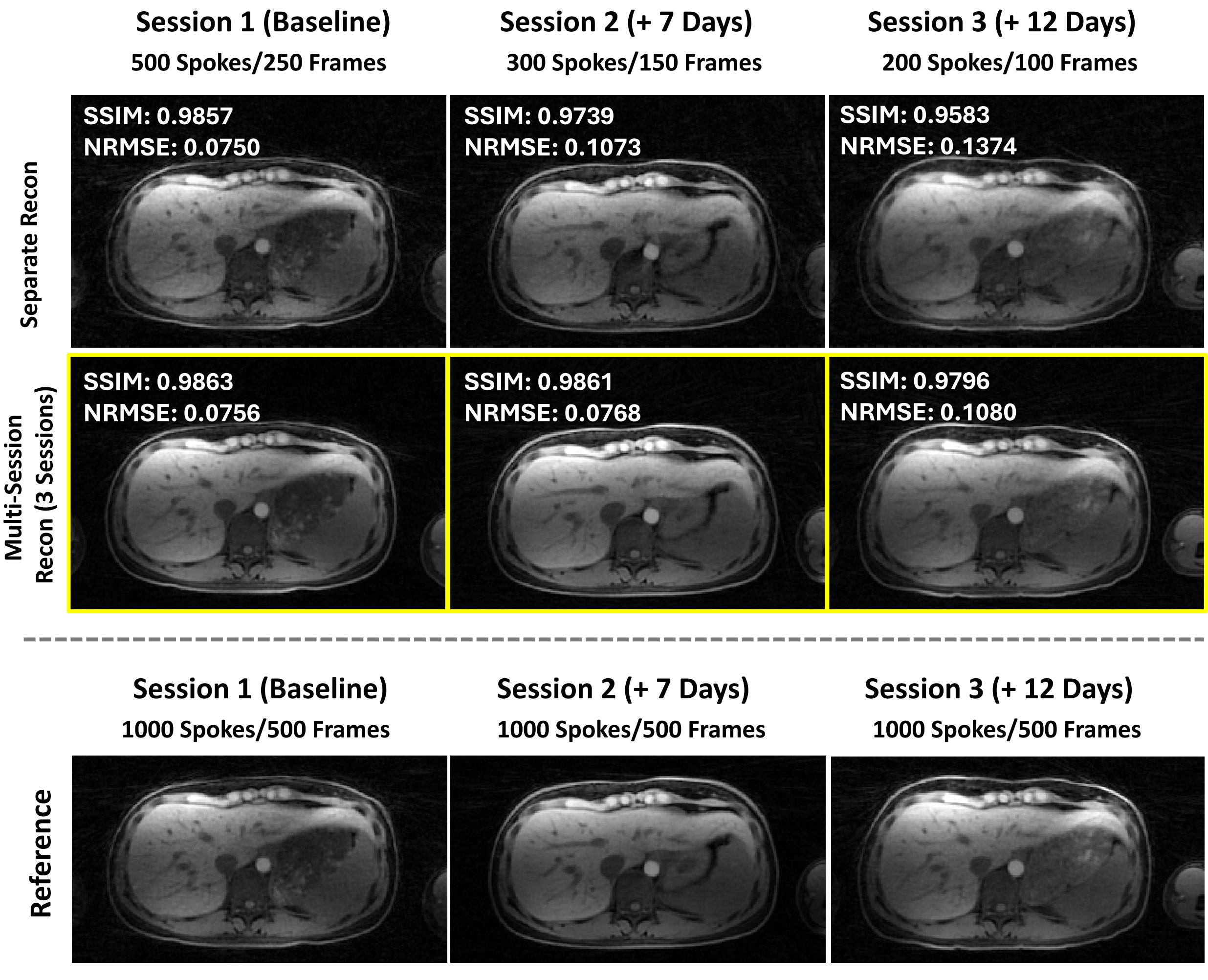

**Figure 2.** Results for experiment 1: added value of subject-specific priors. A snapshot of dynamic images reconstructed using different methods for comparison. 500 radial spokes yielding 250 temporal frames from session 1, 300 spokes yielding 150 temporal frames from session 2, and 200 spokes yielding 100 temporal frames from session 3 are used following the description in **Supplementary Figure S1**. Top row: separate reconstruction, treating each session in isolation. Middle row: joint reconstruction of all three sessions of data concatenated together. Bottom row: The reference dynamic images, shown in the bottom row, are reconstructed using a total of 1000 spokes (500 temporal frames) for each session separately. SSIM and NRMSE values were calculated frame-by-frame for all frames within each session against the corresponding frames in the reference image sets.

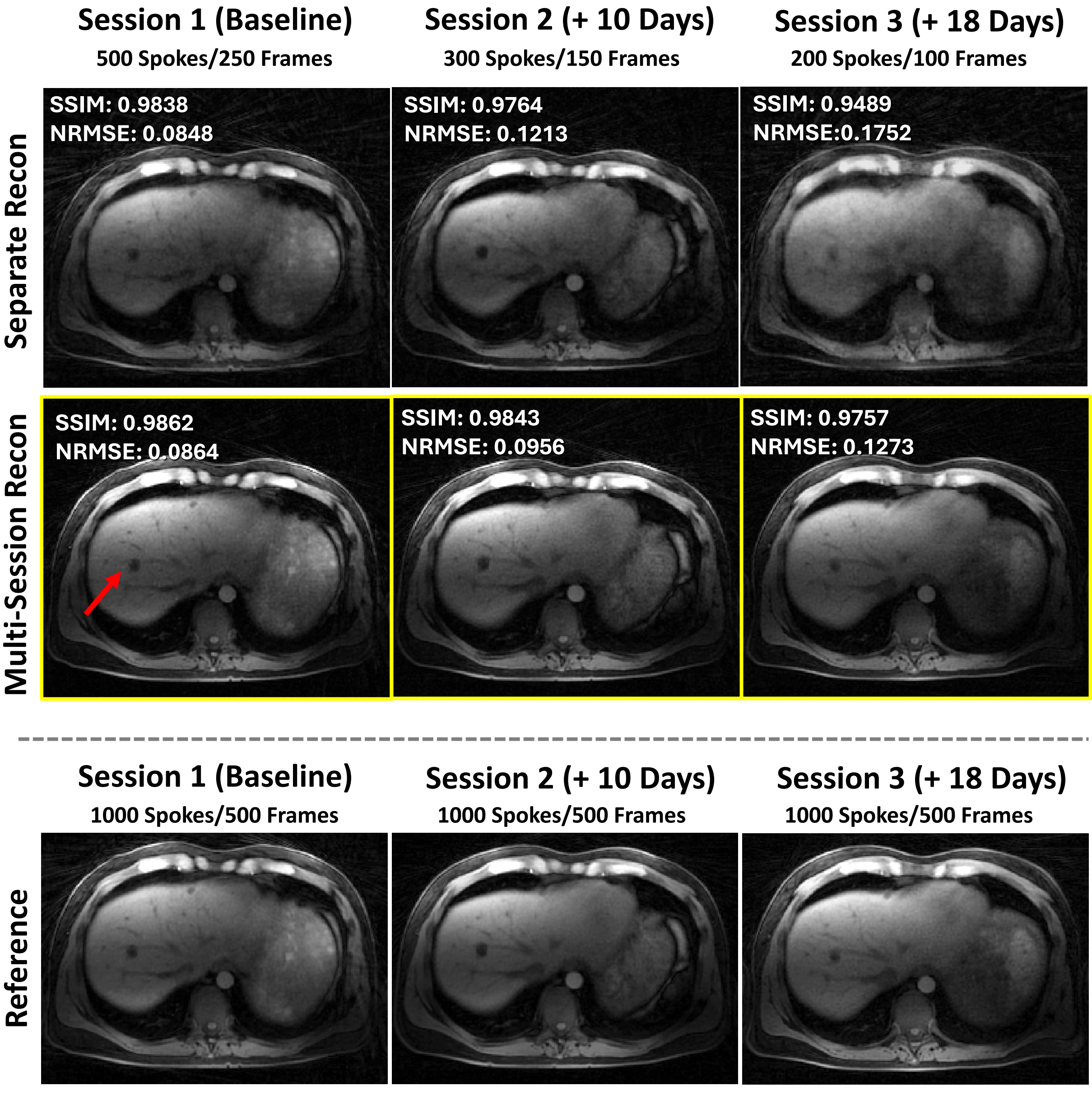

**Figure 3.** Results for experiment 2: preservation of lesions. Longitudinal multi-session reconstruction compared to a separate single-session reconstruction for a subject with a chronic liver lesion. The lesion (red arrow) is largely the same but shows subtle contrast differences across multiple sessions according to the reference images. The multi-session longitudinal 4D GRASP MRI reconstruction recovers these subtle differences with improved image clarity and reduced artifacts as compared with single-session reconstructions.

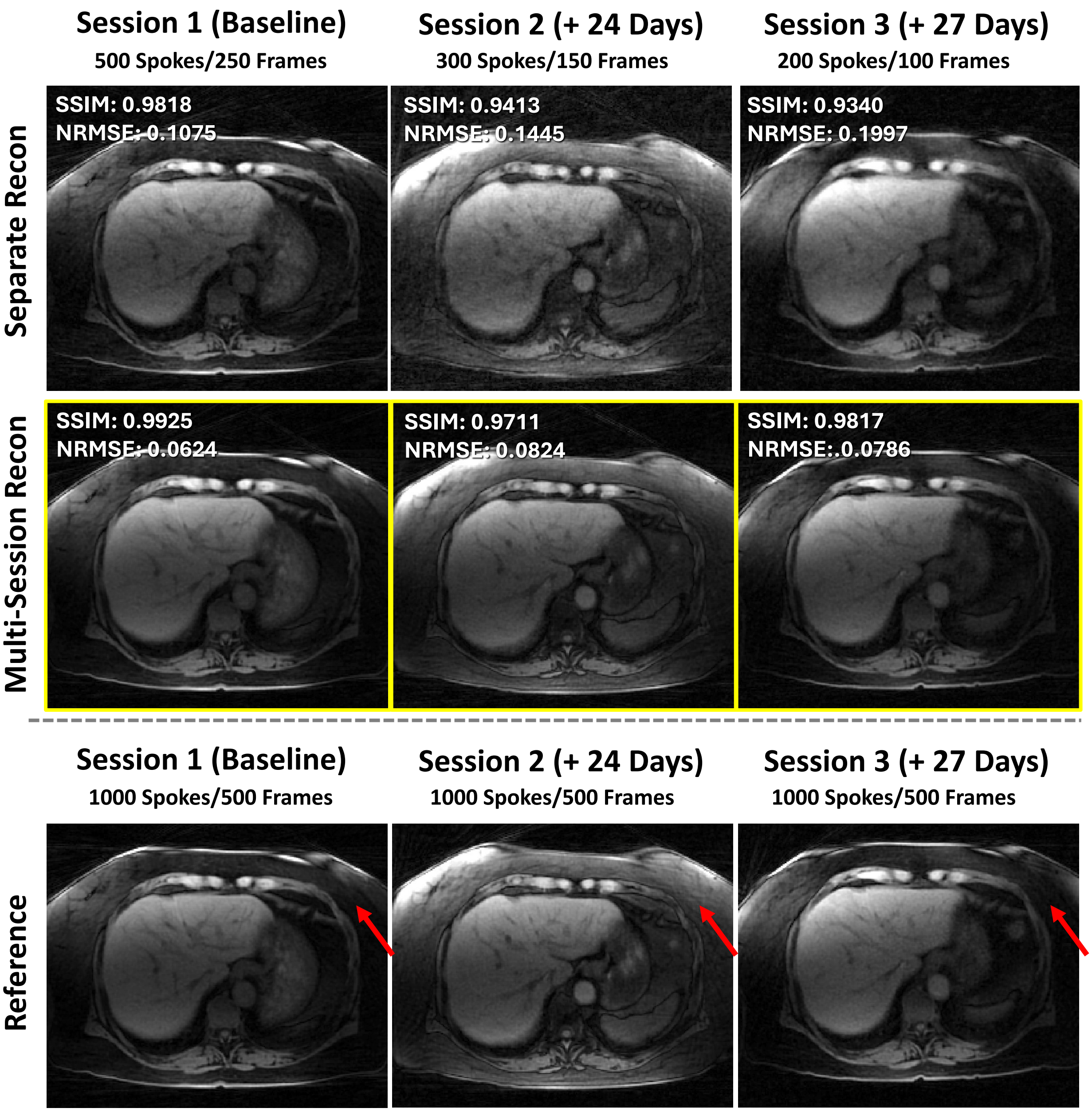

**Figure 4.** Results for experiment 3: robustness to changes across sessions. Longitudinal multi-session reconstruction compared to a separate single-session reconstruction for a subject with different fat signal appearance in distinct sessions. While all three sessions were acquired with fat suppression, the setting for the second session was deliberately sub-optimal, resulting in noticeable residual fat signal (red arrows). Not only does the longitudinal reconstruction improve image quality, but it also does not introduce contrast blurring across sessions.

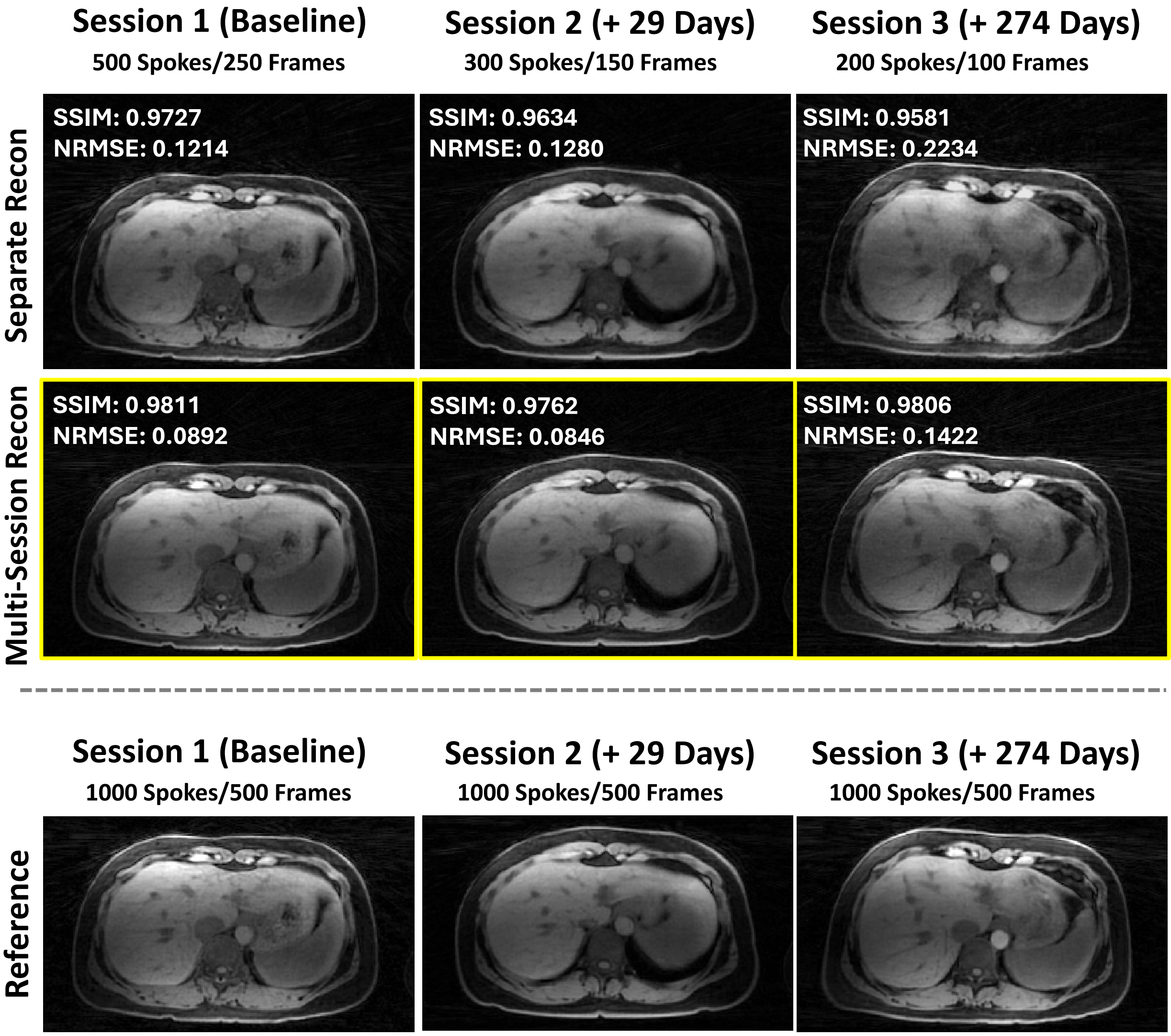

**Figure 5.** Results for experiment 4: robustness to large inter-session gaps. Longitudinal multi-session reconstruction compared to a separate single-session reconstruction for a subject with extended gaps between imaging sessions. The contour of this subject's body changed significantly between sessions, as did the subject's respiration patterns. The longitudinal multi-session reconstruction can still handle such a situation, providing improved image quality as compared with single-session reconstructions, without any leakage from one session to another.

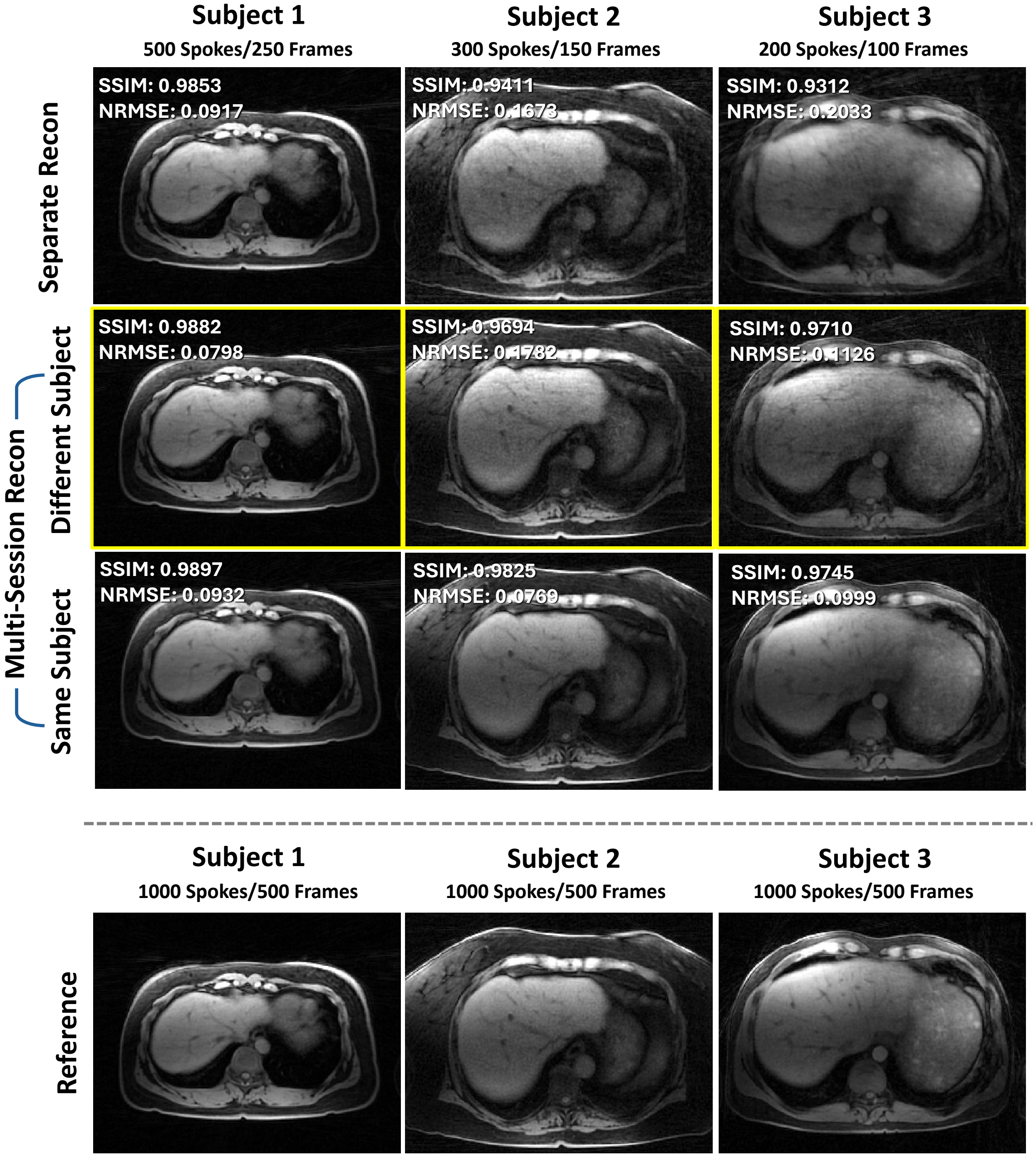

**Figure 6.** Results for experiment 5: subject-specific versus population-based priors. Multi-session reconstruction using data from different subjects as opposed to data from the same subject. For the snapshot images of multi-session reconstruction in the same subject, three sessions of data were acquired on different dates, and only the corresponding sessions are shown and used for image quality metric calculation. Multi-session reconstruction shows improved image quality as compared with separate reconstruction in both cases, but noticeable artifacts appear when concatenating data across different subjects.

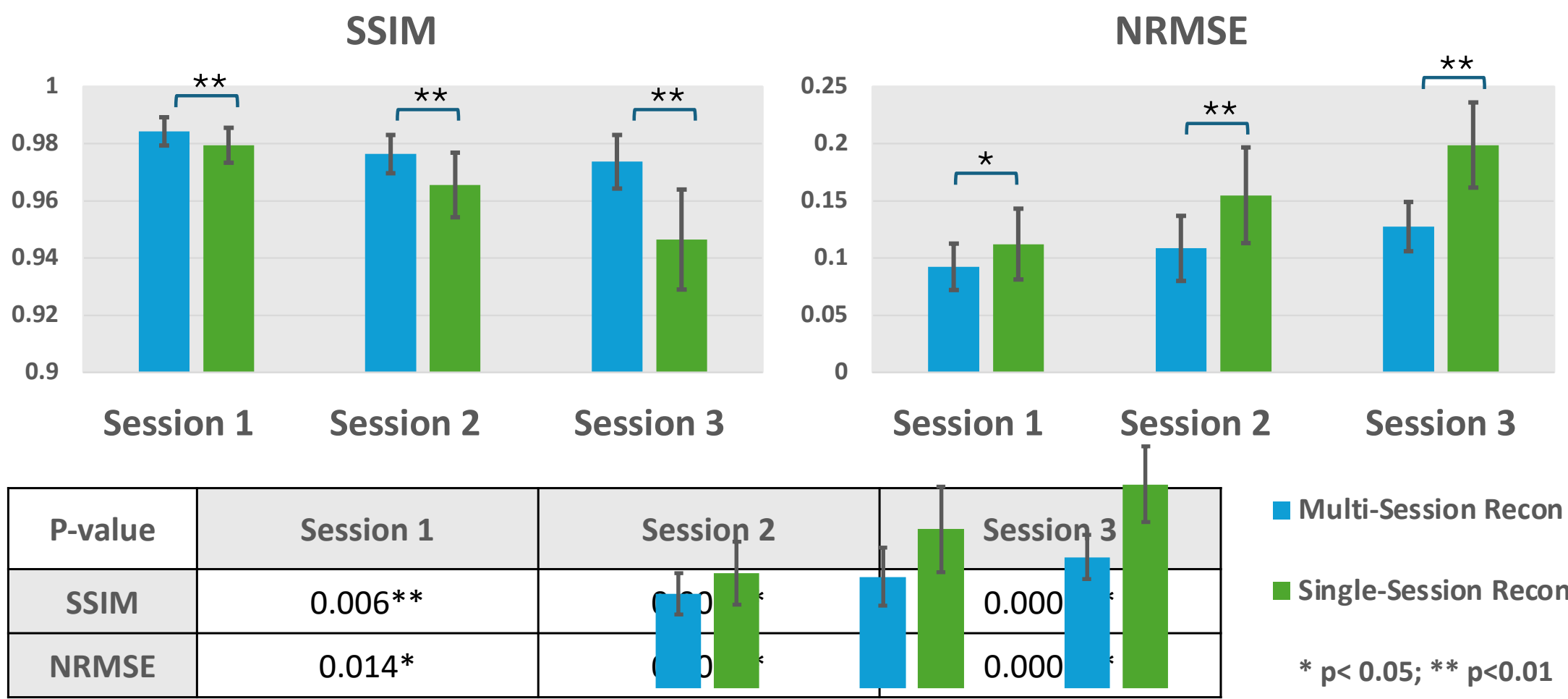

**Figure 7**. Results for experiment 6: Bar plot collecting the SSIM and NRMSE statistics for longitudinal multi-session 4D GRASP MRI reconstruction and separate reconstruction for three sessions, respectively. The error bar indicates the standard deviations. Longitudinal reconstructions show significant improvement in SSIM and NRMSE as compared with single session reconstruction. Asterisks denote statistical significance with *: p<0.05, and **: p<0.01. Detailed p-value is summarized in the table below.

## Supplementary Information

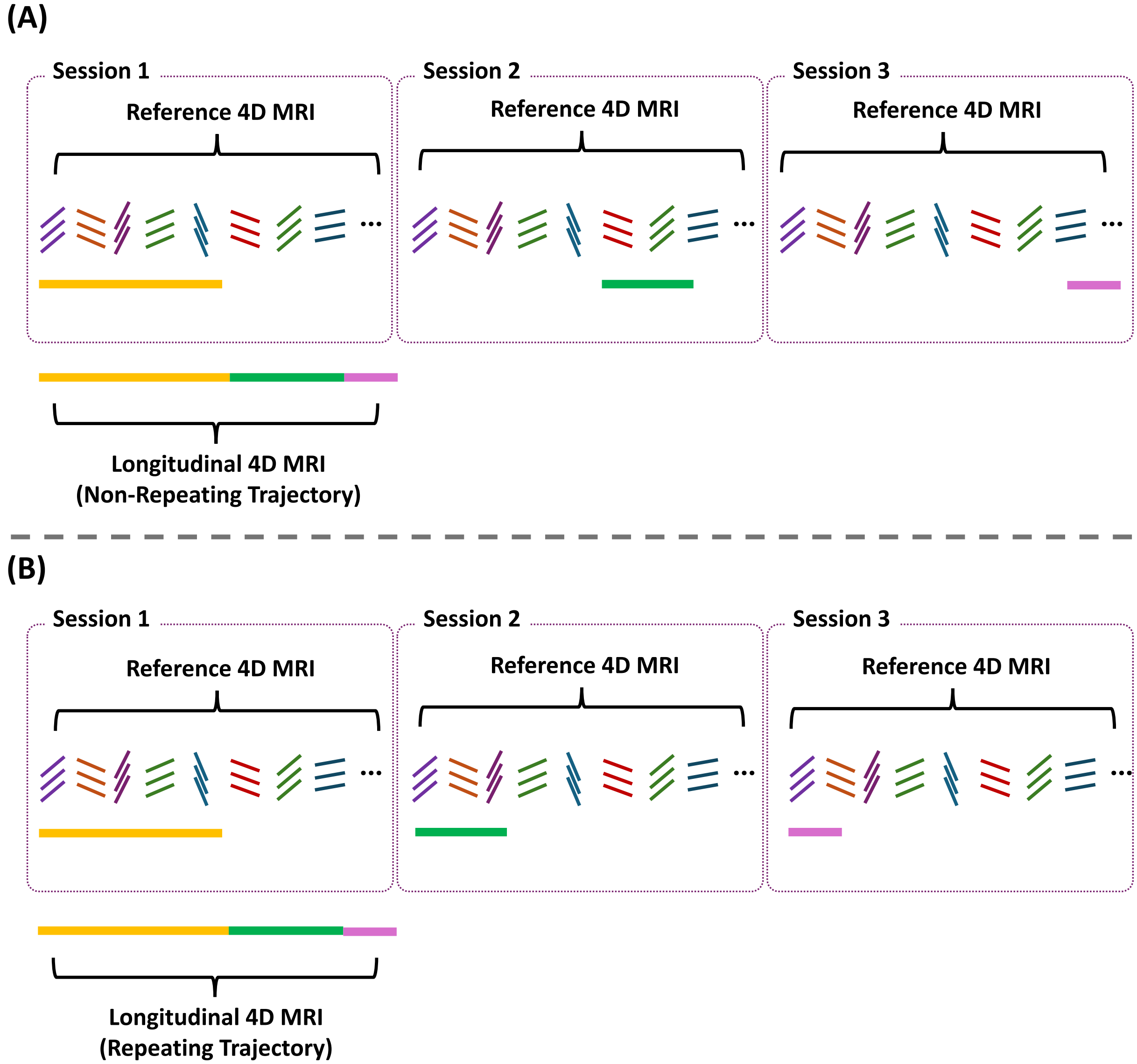

## Supplementary Figure S1

Data concatenation process for longitudinal 4D GRASP MRI reconstruction. (A) In non-repeating trajectory scenario, different sections of the full-length dynamic imaging data are put together to form a single longitudinal dynamic imaging series for the multi-session longitudinal 4D MRI reconstruction. (B) In the case of repeating trajectory, sections are chosen from the beginning the full-length dynamic imaging data and are assembled to form a single longitudinal dynamic imaging series. In this way, the trajectory of the second and the third sessions are subsets of the trajectory of the first session.

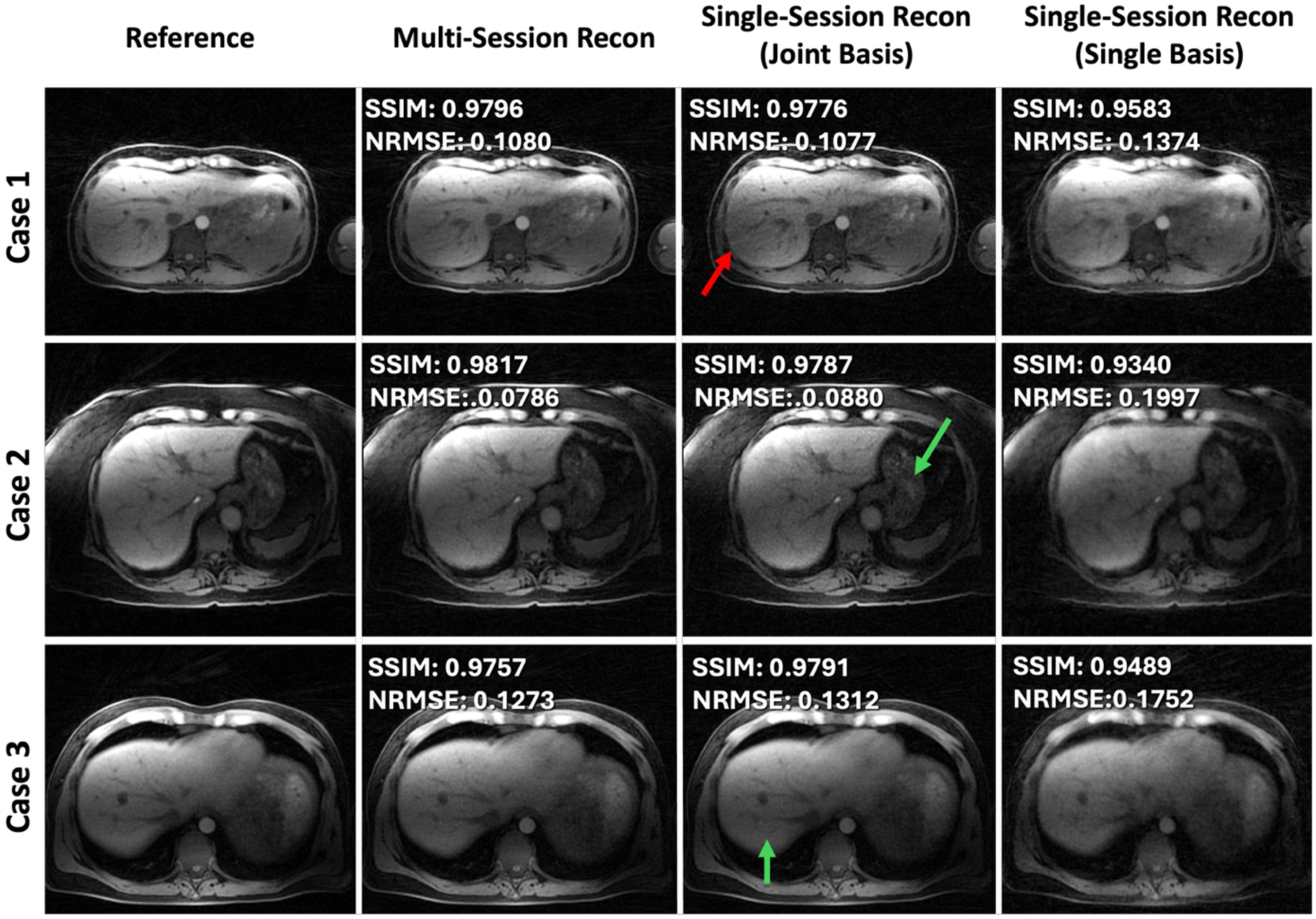

**Supplementary Figure S2**

Comparison of Multi-Session and Single-Session Reconstructions Using Different Subspace Bases for Session 3. The multi-session reconstruction combines imaging data from all three sessions: 500 imaging spokes from session 1, 300 from session 2, and 200 from session 3. The joint basis refers to the subspace basis estimated from the full set of 1000 spokes concatenating all sessions. For single-session reconstruction of session 3, the segment of the joint basis corresponding to session 3 is extracted and applied. As a comparison, a single basis learned solely from the 200-spoke data of session 3 is also used to reconstruct the same session. The reference image is reconstructed using 1000 imaging spokes, consistent with the reference standard used in all other experiments.

**Supplementary Experiment 1: To assess the impact of inter-session alignment on longitudinal 4D GRASP reconstruction.**

Image registration can be applied to all datasets acquired across different imaging sessions to improve alignment for joint multi-session reconstruction, and this step was incorporated by default in all the experiments described above. This experiment aimed to assess whether our reconstruction remains effective without this registration step. To evaluate this, we compared the image quality of longitudinal 4D GRASP reconstructions with and without the pre-alignment step using the same volunteer dataset from Experiment 1. The hypothesis was that image registration would more closely align images from different sessions, thereby increasing temporal sparsity and improving reconstruction quality. However, even in the absence of this alignment step, our reconstruction algorithm is expected to inherently account for inter-session misalignment.

**Supplementary Figure S3** below compares longitudinal 4D GRASP reconstruction with and without a pre-processing step involving rigid registration to improve image alignment across sessions in two different motion phases. Note that without alignment, the images appear different, particularly in the second session, due to imperfect slice-to-slice matching across the three scans. The results indicate that the pre-alignment step improves image quality and reduces residual streaking artifacts, especially in the third imaging session (see red arrows). Although the improvement is not dramatic, it is likely attributable to increased sparsity achieved through better inter-session alignment, as further supported by the quantitative metrics displayed above the images. Nevertheless, good image quality can still be achieved without the pre-alignment step. Note that these metrics were calculated across all dynamic frames, so the numbers are only shown on one panel. Corresponding cine movies for this comparison are provided in **Video 5** (Supplementary Materials).

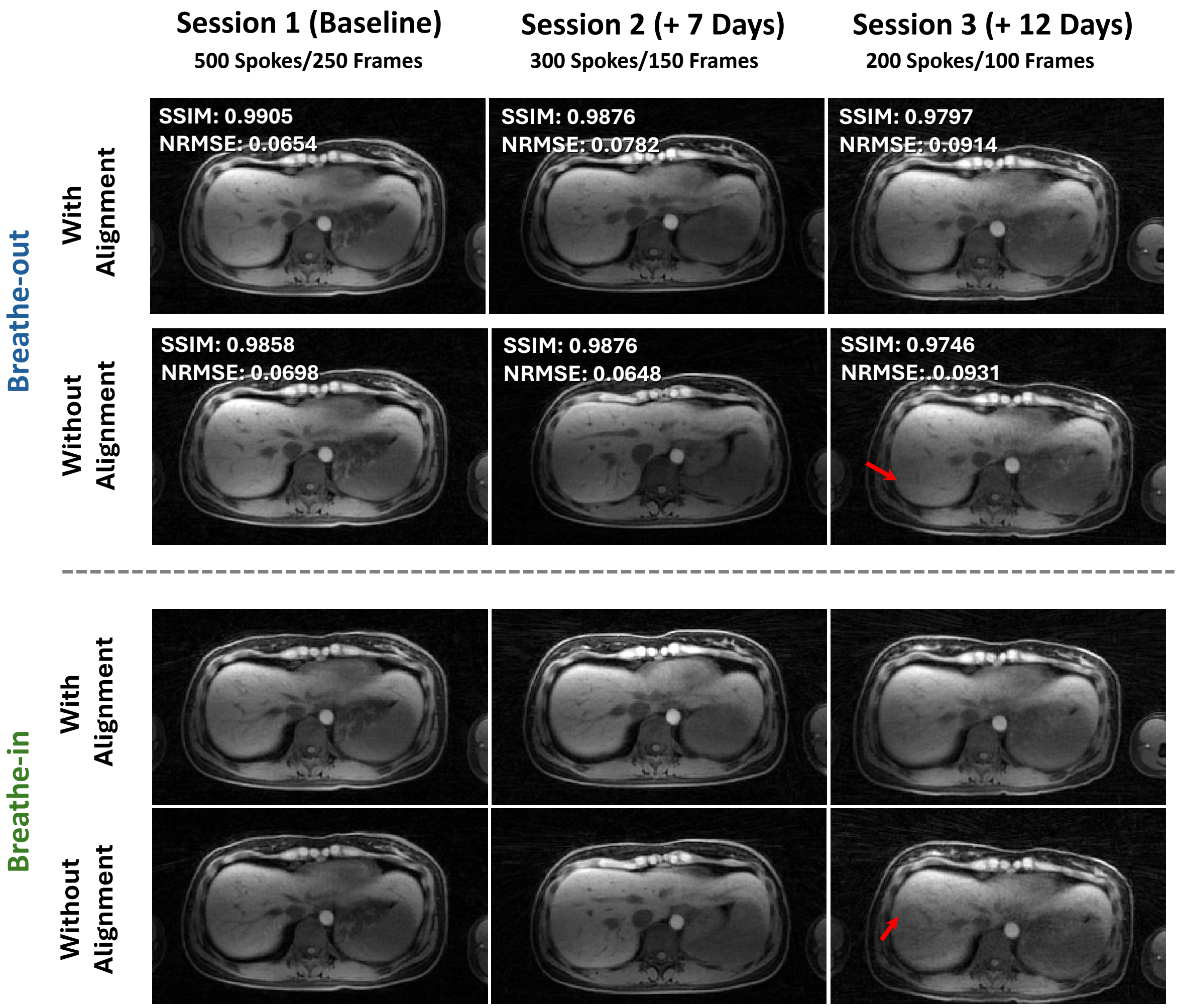

**Supplementary Figure S3.** Comparison of Longitudinal 4D GRASP MRI reconstruction with and without the rigid registration alignment step for two different respiration phases during exhalation and inhalation. Artifacts and signal loss (red arrows) can be seen when alignment is not performed.

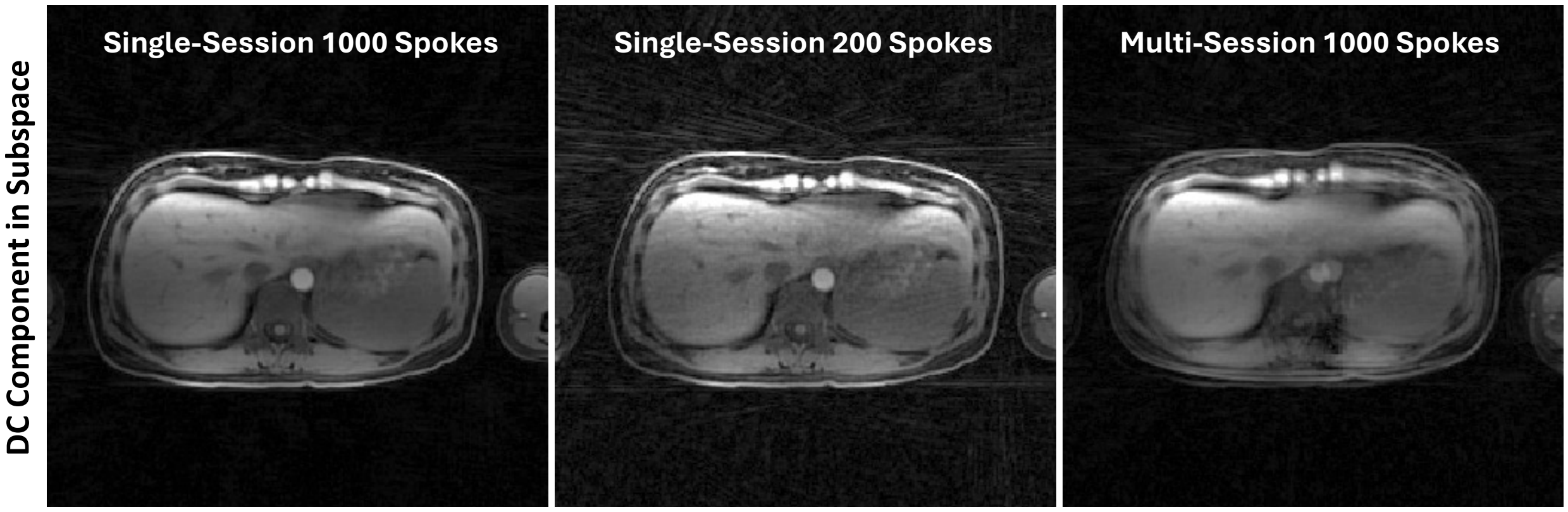

**Supplementary Figure S4**

In subspace reconstruction, the coefficients corresponding to the first basis component are referred to as the DC component. This component captures the most commonly shared information across all temporal frames and typically resembles a motion-averaged image. The quality of the DC component fundamentally influences the fidelity of the final dynamic MRI reconstruction.

This figure shows the DC components derived from: (1) data acquired with 1000 imaging spokes (serving as the reference), (2) data from session 3 using only 200 imaging spokes, and (3) multi-session data combining all three sessions for a total of 1000 imaging spokes. Compared to the reference, the DC component from the 200-spoke dataset exhibits prominent streaking artifacts, both in the background and within the anatomy, due to inadequate k-space coverage. In contrast, the DC component reconstructed from the concatenated multi-session data is free of such artifacts, providing a clean and reliable foundation for high-quality subspace-based 4D GRASP reconstruction.

**Supplementary Experiment 2: To assess whether longitudinal 4D GRASP reconstruction remains effective when the same sampling trajectory is used across imaging sessions.**

In the first experiment, we described how the selection of spokes from the three imaging sessions was designed to ensure a non-repeating sampling trajectory (see **Supplementary Figure S1**). This approach is expected to maximize temporal incoherence to ensure optimal iterative reconstruction with sparsity constraints. In this experiment, we evaluated whether longitudinal 4D GRASP reconstruction would remain effective when spokes with overlapping rotation angles (referred to as repeating sampling trajectory) were selected. The experiment was performed on the subject with a liver lesion (used in Experiment 2). The selection of repeating spokes for different sessions is also shown in **Supplementary Figure S1.** Specifically, the 500 spokes from the first session remained unchanged, while the 300 spokes from the second session shared the same trajectory as the first 300 spokes from the first session. Similarly, the 200 spokes from the third session shared the same trajectory as the first 200 spokes from both the first and second sessions. This results in overlapping sampling patterns across imaging sessions, which is expected to reduce temporal incoherence for iterative reconstruction.

**Supplementary Figure S5** compares longitudinal 4D GRASP reconstruction using non-repeating versus repeating sampling trajectories across different imaging sessions. While the images reconstructed with non-repeating trajectories appear slightly better in the x-t plots, the overall image quality between the two methods is visually comparable and the observed differences are subtle. However, based on theoretical advantages in maintaining temporal incoherence and quantitative metric, we chose to use non-repeating trajectories in our longitudinal reconstruction in all previous experiments. This comparison was also performed in all subjects, and the corresponding results are presented in the **Supplementary Figure S6**.

**Impact on Sampling Trajectories Across Imaging Sessions**

In Supplementary Experiment 2, we compared longitudinal 4D GRASP reconstruction using non-repeating versus repeating sampling trajectories across sessions. Although non-repeating trajectories are theoretically preferred and demonstrated slightly better image quality (see **Supplementary Figure S5**), the difference was not significant. One reason for this robustness is that concatenating data across multiple sessions still improves the estimation of the temporal basis function, as shown in Supplementary Figure S2 (Supplementary Materials). This suggests that while non-repeating trajectories maximize temporal incoherence for optimal reconstruction, our method remains flexible with respect to sampling patterns and can potentially accommodate different types of sampling strategies.

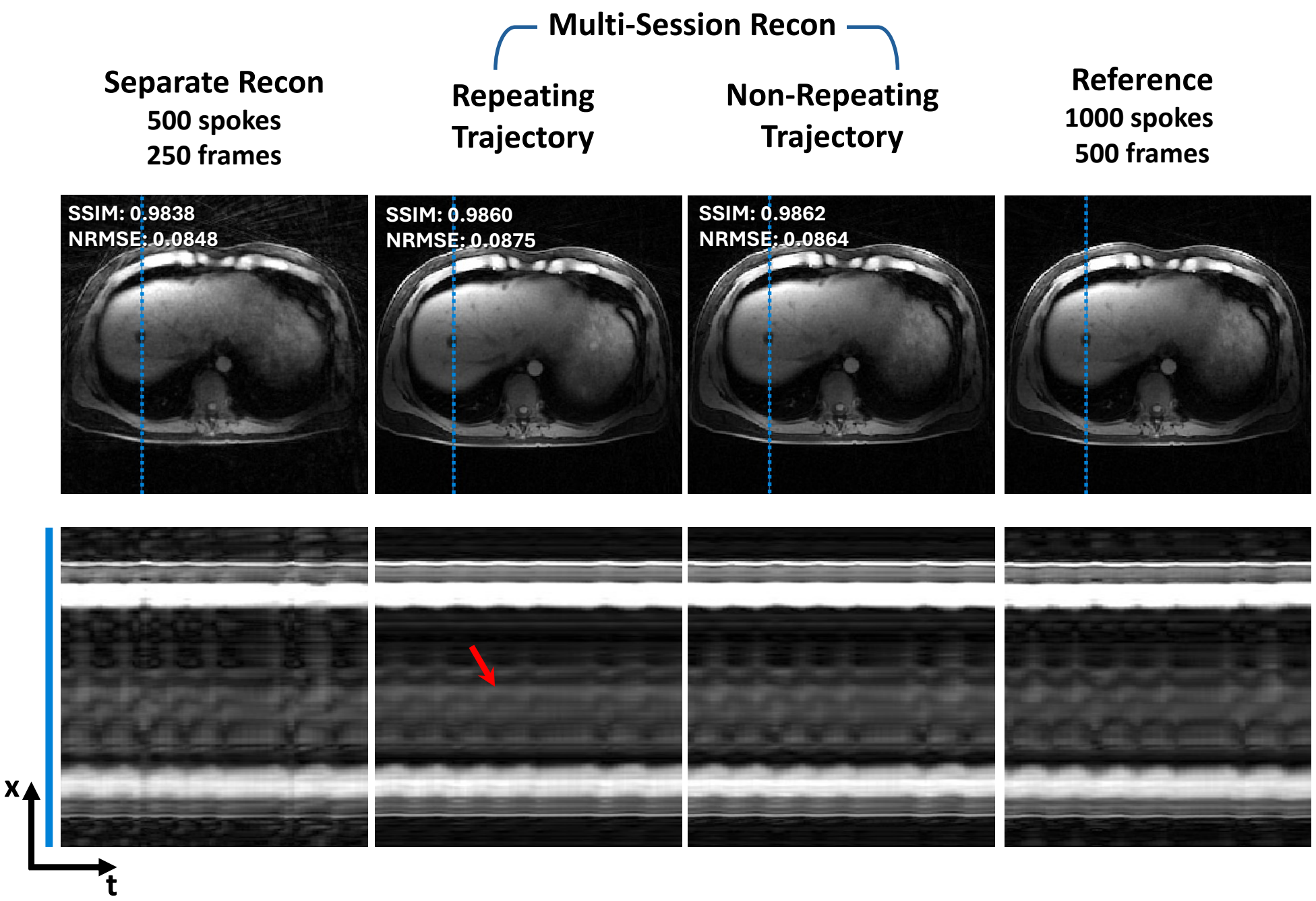

**Supplementary Figure S5.** Longitudinal multi-session reconstruction using repeating and non-repeating k-space trajectory for each session compared to the separate reconstruction. The images are showing the same dynamic frames for different reconstructions. The x-t plots correspond to the temporal dynamics of the blue dashed profile line labeled in the images. While the reconstructed image looks similar when three sessions are acquired with repeating and non-repeating trajectory, the x-t plot reveals slight temporal blurring (red arrow) when multi-session data are acquired with repeating trajectory.

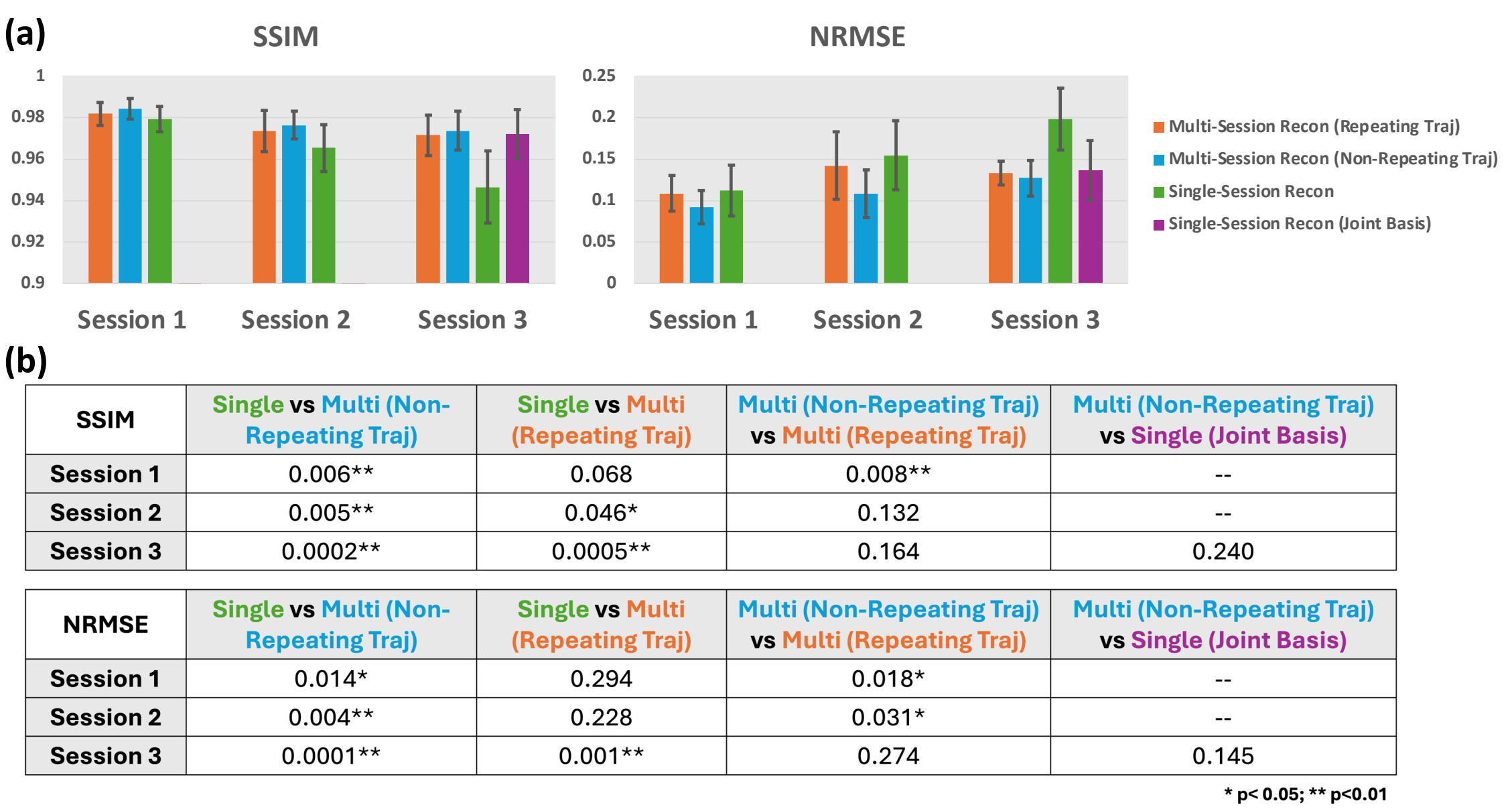

| SSIM | Single vs Multi (Non-Repeating Traj) | Single vs Multi (Repeating Traj) | Multi (Non-Repeating Traj) vs Multi (Repeating Traj) | Multi (Non-Repeating Traj) vs Single (Joint Basis) |
|---|---|---|---|---|
| **Session 1** | 0.006** | 0.068 | 0.008** | -- |
| **Session 2** | 0.005** | 0.046* | 0.132 | -- |
| **Session 3** | 0.0002** | 0.0005** | 0.164 | 0.240 |

| NRMSE | Single vs Multi (Non-Repeating Traj) | Single vs Multi (Repeating Traj) | Multi (Non-Repeating Traj) vs Multi (Repeating Traj) | Multi (Non-Repeating Traj) vs Single (Joint Basis) |
|---|---|---|---|---|
| **Session 1** | 0.014* | 0.294 | 0.018* | -- |
| **Session 2** | 0.004** | 0.228 | 0.031* | -- |
| **Session 3** | 0.0001** | 0.001** | 0.274 | 0.145 |

* p< 0.05; ** p<0.01

**Supplementary Figure S6.** (a) Bar plot collecting the SSIM and NRMSE statistics for longitudinal multi-session 4D GRASP MRI reconstruction with non-repeating trajectory, multi-session reconstruction with repeating trajectory, and separate reconstruction for three sessions, respectively. The error bar indicates the standard deviations. Additionally, the purple error bars represent single-session reconstruction using only single session data, but the segmented basis estimated from longitudinal data. This experiment was performed only on session 3, which contains the least amount of data, thereby making the effect of improved basis estimation more apparent. (b) Summary of p-values of t-test comparing different reconstruction strategies. Longitudinal reconstructions show significant improvement in SSIM and NRMSE as compared with single session reconstruction. Longitudinal reconstructions with repeating or non-repeating trajectory show no significant difference for session 2 and 3 and a slight significant difference for session 1. Asterisks denote statistical significance with *: p<0.05, and **: p<0.01.